\documentclass[review,12pt,3p]{elsarticle}

\usepackage{amssymb, amsmath, siunitx,amsfonts,mathrsfs}
\usepackage{lineno,setspace}
\usepackage{commath,upgreek,physics}

\journal{Journal of Colloid and Interface Science}

\begin{document}

\begin{frontmatter}

\title{From coffee stains to uniform deposits: significance of the contact-line mobility}

\author[1,2]{Aleksander Matav\v z\corref{cor1}\fnref{fn1}}
\author[1]{Ur\v{s}a Ur\v{s}i\v{c}}
\author[1]{Jaka Mo\v{c}ivnik}
\author[1]{Dmitry Richter\fnref{fn2}}
\author[1,3,4]{Matja\v{z} Humar}
\author[1,3]{Simon \v{C}opar}
\author[1,2]{Barbara Mali\v c}
\author[1,2]{Vid Bobnar}

\address[1]{Jo\v{z}ef Stefan Institute, Jamova 39, SI-1000 Ljubljana, Slovenia}
\address[2]{Jo\v{z}ef Stefan International Postgraduate School, Jamova 39, SI-1000 Ljubljana, Slovenia}
\address[3]{Faculty of Mathematics and Physics, University of Ljubljana, Jadranska 19, SI-1000 Ljubljana,
Slovenia}
\address[4]{CENN Nanocenter, Jamova 39, SI-1000 Ljubljana, Slovenia}

\cortext[cor1]{Corresponding author. Email: aleksander.matavz@ijs.si   }
\fntext[fn1]{Present address: KU Leuven, Celestijnenlaan 200F, 3000 Leuven, Belgium}
\fntext[fn2]{Present address: Wellman Center for Photomedicine, Harvard Medical School and Massachusetts General Hospital, Boston, MA 02114, USA}

\begin{abstract}
\noindent \textit{Hypothesis}\\
Contact-line motion upon drying of sessile droplet strongly affects the solute transport and solvent evaporation profile. Hence, it should have a strong impact on the deposit formation and might be responsible for volcano-like, dome-like and flat deposit morphologies.

\noindent \textit{Experiments}\\
A method based on a thin-film interference was used to track the drop height profile and contact line motion during the drying. A diverse set of drying scenarios was obtained by using inks with different solvent compositions and by adjusting the substrate wetting properties. The experimental data was compared to the predictions of phenomenological model.

\noindent \textit{Findings}\\
We highlight the essential role of contact-line mobility on the deposit morphology of solution-based inks. A pinned contact line produces exclusively ring-like deposits under normal conditions. On the contrary, drops with a mobile contact line can produce ring-, flat- or dome-like morphology. The developed phenomenological model shows that the deposit morphology depends on solvent evaporation profile, evolution of the drop radius relative to its contact angle, and the ratio between initial and maximal (gelling) solute concentration. These parameters can be adjusted by the ink solvent composition and substrate wetting behaviour, which provides a way for deposition of uniform and flat deposits via inkjet printing.

\end{abstract}

\begin{graphicalabstract}
\includegraphics{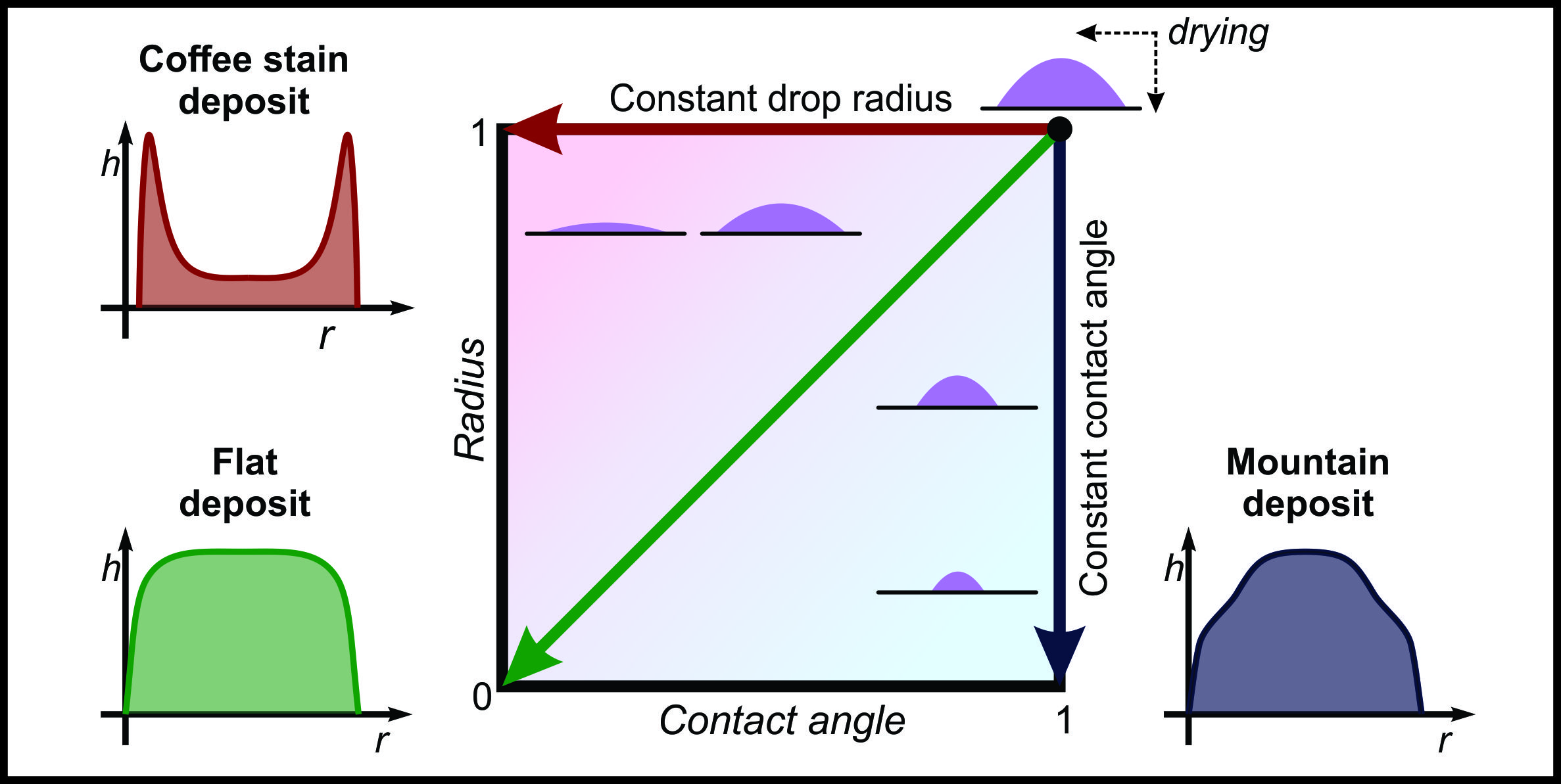}
\end{graphicalabstract}


\end{frontmatter}

\section*{Introduction}
When a drop containing a solute dries on a surface, it leaves behind a solid deposit. Deposits may exhibit a rich diversity of morphologies. The most common is the so-called ``coffee stain'' --- a ring-like morphology with most of the solute concentrated along the perimeter of the initial drop. The formation of a ``coffee stain'' deposit is not limited only to spills of coffee but commonly occurs in technologically relevant systems, such as particle dispersion inks and polymer/sol-gel solutions in aqueous or organic solvents. At present, the inability to control the deposit morphology is a significant bottleneck in printing technologies where uniform deposits are required. 

Deegan et al.\cite{Deegan1997} showed that the solute accumulation at the edge occurs due to an outward capillary flow, which appears due to the contact line pinning and/or edge-enhanced evaporation. Many strategies have been proposed to overcome the deposit accumulation at the edge --- by either influencing the liquid mass transfer (LMT) or the solute mass transfer (SMT). The most prominent methods to tailor the LMT rely on inducing the surface-tension-driven flows,\cite{Hu2006} modifying the evaporation field,\cite{Harris2007} and pinning/unpinning the contact line.\cite{Li2013} On the other hand, the available parameters to control the SMT depend on the physical form of the solute. For particle-dispersion inks, the SMT depends on the particle size and shape,\cite{Yunker2011} jamming at the contact line wedge,\cite{Dan2018} and the adhesion at the liquid--vapour or liquid--solid interfaces.\cite{Li2016} For solution inks, such as polymer or sol-gel inks discussed here, the SMT is influenced mainly by the point of sol-to-gel transition, \textit{i.e.} rapid increase in the viscosity.

A strong cross-correlation between LMT and SMT exists in drying drops; hence, direct deduction of links between drying parameters and final deposit morphology can be difficult and ambiguous. In this aspect, modelling offers a valuable insight into the most important parameters in deposit formation. The majority of available models assume a permanently pinned contact line and can successfully reproduce the formation of ring-like deposits. However, under realistic experimental conditions, the models with the permanently pinned contact line cannot reproduce the formation of deposits with different morphology, such as dome-shaped or flat deposits.\cite{Eales2015,Eagles2016} To explain different morphologies, surface-tension-driven flow (Marangoni flow) was introduced into the models;\cite{Diddens2017} however, there is often a disagreement between the observed and calculated deposit morphology.\cite{Babatunde2013,Eagles2016} More recently, the mobility of the contact line has been considered to explain the formation of dot-like,\cite{Li2013}, volcano-like,\cite{Kajiya2009} and dome- or mountain-like deposits.\cite{Matavz2018} Brown et al.\cite{Brown2014} described the mountain-like deposit formation by considering drying at a constant contact angle. Man et al.\cite{Doi2016} introduced a contact line friction to reproduce different deposit shapes. These pioneering studies imply that the mobility of the contact line may indeed be a crucial parameter for the deposit morphology. Yet, several aspects remain unclear at present.

This paper highlights the effects of the contact line mobility on the deposit formation in inkjet-printed drops. First, we derive a phenomenological model for thin homogeneous drops with a mobile contact line. The unique input parameter of the model is the evolution of the drop radius relative to its contact angle, which can be readily extracted from the drop's height profiles acquired during drying. The model successfully reproduces ``coffee stain'', flat and mountain-like deposit morphology, depending on the contact-line motion. The model predictions are then compared to experimental data. Diverse drying situations are obtained by using two inks with different solvent composition on surfaces with tailored wettability. The results show that deposit morphology depends directly on the contact-line mobility and the evaporation profile. In the last part, we discuss the morphology of inkjet-printed square patterns as a function of the contact-line mobility and elucidate the effect of the drop spacing on the shape of printed features.

\section*{Materials and Methods}
\subsection*{Inks}
The inks used in the study consisted of lanthanum nitrate hydrate (Sigma Aldrich, 99.9\%) and nickel acetate hydrate (Alfa Aesar, 99\%) in 1:1 molar ratio ($c=0.1$~\si{\mole\per\liter}, $\phi=2.8$~\si{\gram\per\gram}) dissolved in 2-ethoxyethanol (Alfa Aesar, 99\%) and combined with ethylene glycol (EG, Riedel-de Haen, 99.5\%) or/and ethanolamine (EA, Sigma Aldrich, 99.5\%). The volume ratio of 2-ethoxyethanol and EG or EA was 20:80. EA/EG ink reported in \figurename~\ref{Fig6}e was synthesised by combining the solution of the metal salts in 2-ethoxyethanol with ethanolamine and ethylene glycol in 20:64:16 volume ratio. 

\subsection*{Substrates and regulation of wetting}
Silicon wafers were cleaned by a sequential ultrasonic cleaning in acetone, 2-propanol and deionized water. The substrates were dried by blowing with nitrogen gas and heating at 350~\si{\celsius} for 10 minutes. The substrates were then coated by about 10~\si{\nano\meter} thick nanotextured poly(methyl methacrylate)/poly(styrene) layer (PMMA/PS). The details of polymeric layer synthesis and properties are reported in \cite{Matavz2019,Matavz2021}. The polymer-coated substrates were dried on a hotplate at 200~\si{\celsius} for 10~\si{\minute} immediately after spin coating. To achieve distinct wetting properties, the polymeric layers were treated by UV/ozone (Ossila Ozone cleaner) for different times and optionally post-annealed (PA) at 100~\si{\celsius}. The following procedures were used to achieve the specific drying regimes: 150~\si{\second} UV/ozone + 10~\si{\second} PA (pinned EG ink, \figurename~\ref{Fig2}), 150~\si{\second} UV/ozone (unpinned EG ink, Drop I in \figurename~\ref{Fig3}), 240~\si{\second} UV/ozone (unpinned EG ink, Drop II in \figurename~\ref{Fig3}), 30~\si{\second} UV/ozone (unpinned EA ink, Drop II in \figurename~\ref{Fig5}), 150~\si{\second} UV/ozone + 10~\si{\second} PA (unpinned EA ink, Drop I in \figurename~\ref{Fig5}).

\subsection*{Drop generation and drying experiments}
About 2~\si{\nano\liter} sessile droplets of ink were deposited on a substrate at ambient conditions using a piezoelectric dispensing unit (MicroFab MJ-AL-01-50-6MX) with an orifice diameter of 50~$\upmu$m. 30 individual droplets were burst-jetted with a frequency of \SI{1}{\kilo\hertz} to reach the target volume. After deposition of a droplet, the sample was transferred into a custom-designed environmental chamber with a hotplate. The chamber was mounted in an upright optical microscope (Zeiss Axio Imager.A1m) configured in reflection geometry (SI Fig. S6). White light was passed through a narrow bandwidth filter (CWL$=$632.8 nm, FWHM$=$10 nm) for illumination. After positioning the sample, the hotplate temperature was increased from \SI{24}{\celsius} to \SI{50}{\celsius} with a heating rate of \SI{150}{\kelvin\per\minute} and the images were acquired in \SI{1}{\second} interval. The relative humidity in the environmental chamber was maintained at 40\% unless specified otherwise. 

The initial height profile EG-based ink drop with pinned contact line was determined by circle-segment fit to the drop cross-section recorded using a Krüss DSA20E tensiometer. The height profiles of other droplets were reconstructed using a thin-film interference method (\figurename~\ref{Fig1add}). The interference pattern occurs due to the phase changes of the light wave reflected at the upper boundary (liquid-–air interface) and lower boundary (liquid–-substrate). The phase difference is due to different path lengths of both waves with the condition for constructive interference of reflected light
\begin{equation}
2 n_{\mathrm{ink}} h \cos{\theta_i} = m \lambda,
\label{constructive}
\end{equation}
and the condition for destructive interference
\begin{equation}
2 n_{\mathrm{ink}} h \cos{\theta_i} = \left( m-\frac{1}{2} \right) \lambda.
\label{destructive}
\end{equation}
$n_{\mathrm{ink}}$ is the refractive index of the ink, $h$ is the local height, $\theta_i$ is the angle of incidence of the light wave, $m$ is the integer corresponding to the interference order, and $\lambda$ is the wavelength of the light. The drop height profile is then reconstructed by detection of intensity maxima and counting their order from the periphery toward the centre (\figurename~\ref{Fig1add}b). This is performed under the assumption that the thickness of the drop continuously increases from the edge to the centre. This might not be always true as the drop can collapse in the centre of the drop; however, such anomaly is identifiable by tracking the time progression of the interference fringes during the drying process (normally, the fringes appear to move toward the centre of the drop upon drying). The refractive index of the inks increased less than 3\% when solute concentration was increased from 0 to 22~\si{\gram\per\gram}; hence, the variations in $n$ were considered irrelevant in the droplet height reconstruction. In addition, the deposit that formed at the contact line was in a form of a gel and no light-scattering particles were observed that would distort the interference image. The first derivative of height with respect to the radius was used to track the contact line motion and extract the $R(\theta)$ trajectory for each drop.

The detailed height profiles presented in \figurename~\ref{Fig2}c were determined by measuring the reflectance spectra at 0.78~$\upmu$m pitch using Andor Shamrock SR-500i-D1-R spectrometer equipped with Andor Newton camera. The spectra were measured from 520.2--684.4~nm with a resolution of 0.1~nm. The local height of the drop was determined by fitting the reflectance spectra using an idealized model consisting of a three-media structure (air, film, substrate) with a transparent film on a non-adsorbing semi-infinite substrate. For this experiment, the surface-modified glass (Corning XG Eagle) was used as a substrate and was prepared in the same way as silicon wafers (see Substrates and regulation of wetting). The temperature during the drying was maintained at 40~\si{\celsius} using an Okolab H301-K frame and control unit.

\subsection*{Inkjet printing of a pattern}
Square patterns (0.8$\times$0.8~\si{\milli\meter^2}) of different inks were printed using a Fujifilm Dimatix DMP 2831 inkjet printer equipped with a 10~\si{\pico\liter} cartridge. The printed structures were first dried at 80~\si{\celsius} and afterwards heated at 150~\si{\celsius} and 250~\si{\celsius}. All thickness profiles were recorded using a Bruker Dektak XT contact profilometer.

\section*{Results and discussion}
\subsection*{1. Phenomenological model of drying drops}
\begin{figure*}[t!]
\centering
\includegraphics[clip,angle=0,width=14cm]{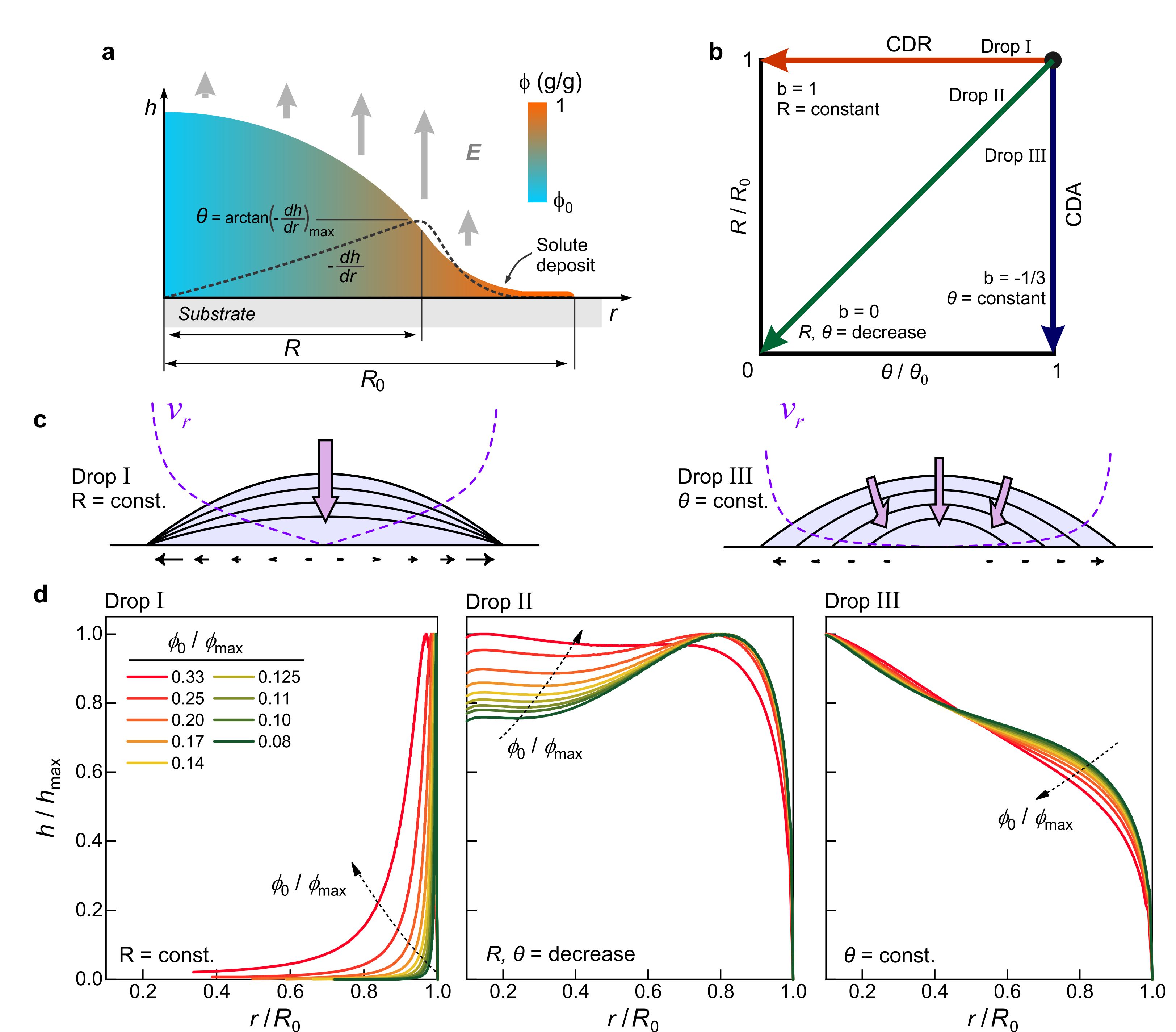}
\caption{\textbf{Simulations of deposit formation for drops with different trajectories in $R(\theta)$ diagram.}  \textbf{a,} Illustration of a cross-section of a drying drop with a receding contact line. The drop contains solute which gets deposited at contact line and generates a typical S-shaped height profile (SI Section 1). Dotted line shows a negative derivative of the height profile, which is used in experiments to extract the position of the drop radius, $R$, and contact angle, $\theta=\arctan(-\mathrm{d}h/\mathrm{d}r)_\mathrm{max}$ during drying. \textbf{b,} $R(\theta)$ diagram showing example of three possible trajectories to dry a drop: (i) only contact angle decreases (CDR), (ii) both contact angle and radius decrease, and (iii) only radius decreases. \textbf{c,} Sketch of the shape evolution and velocity profile of the CDR (i) and CDA (iii) drying regimes. Pinned contact line results in stronger outward flow, leading to the ``coffee stain'' effect. \textbf{d,} The simulated deposit morphology for drop trajectories I, II and III at different initial-to-critical solute concentration.}
\label{Fig1}
\end{figure*}

We focus on drops that are small (with radius $R<1$~\si{\milli\meter}), thin (with thickness $h \ll R$) and with a small contact angle ($\theta<45$\si{\degree})\cite{Popov2005}, \figurename~\ref{Fig1}a. For such drops, the surface tension dominates over the gravitational forces---resulting in a small Bond number, $\mathrm{Bo}<0.05$---and the drops assume a spherical cap shape. Larger drops might follow a different drying mechanism and would require consideration of the scaling relations.\cite{Kaplan2015}

To show which mechanisms underlie the differences in deposit morphology, we construct a minimal model of a droplet shape evolution that does not require modeling of the complex situation at the trailing edge. For thin drops, vertical flow components and concentration gradients can be neglected and the lubrication approximation applies, which yields the continuity equation for the velocity field $\vec{v}$ and thickness $h$, both depending on the radial coordinate $0<r<R$,
\begin{equation}
  \pd{h}{t}+\nabla\cdot (h\vec{v})=E(r)
  \label{eq2}
\end{equation}
with the evaporation profile (also called evaporative flux), $E(r)$. We assume the surface tension ensures that the drop shape is always a spherical cap  approximated by a parabolic profile,  $h(r)=\tfrac{R^2-r^2}{2R}\theta$, where $\theta$ is the contact angle in small angle approximation. Then, $E(r)$ is obtained by solving the Laplace equation in the limit of small contact angles,
\begin{equation}
  E(r)=-\frac{D\Delta c}{\rho} \frac{2}{\pi} \frac{1}{\sqrt{R^2-r^2}}
  \label{eq:evaporation}
\end{equation}
with the vapor diffusivity, $D$, the difference between saturated and ambient vapour concentration, $\Delta c$, and vapor density, $\rho$. In this work we discuss solution inks; hence, the solute transport is directly coupled to the radial flow velocity of the liquid, $v_r(r,t)$, which depends on the evolution of the droplet shape. Deegan and Popov have derived the solution for a droplet with a pinned edge ($R=\text{const.}$).\cite{Deegan1997,Popov2005} As the effect of contact line mobility on the final deposit morphology is the central point of our research, we employ a more general model that can account for any polytropic (power law) change of the radius $R$ and contact angle $\theta$:
\begin{equation}
    \frac{{\rm d}R}{{\rm d}\theta}=\frac{1-b}{1+3b} \frac{R}{\theta}
\end{equation}
with parameter $b$ chosen in order to simplify the solution for radial velocity profile, which solves the continuity equation:
\begin{equation}
v_r=\frac{D\Delta c}{\rho} \frac{4}{\pi}\frac{1}{r\theta}\left(\frac{R}{\sqrt{R^2-r^2}}-\frac{R^2-br^2}{R^2}\right).
\label{RadialFlow}
\end{equation}

This solution includes the pinned drop (constant drop radius, CDR) at $b=1$ and constant contact angle (CDA) drop at $b=-\tfrac13$. In a real situation, the polytropic parameter $b$ can change during evaporation, for example, at the pinning--unpinning transition. The $R(\theta)$ trajectory, which is purely geometrical and can be measured, contains almost all information about the evaporation progress of a droplet and shows whether a constant $b$ model is justified or not.

The velocity profile, Equation~\ref{RadialFlow} (see \figurename~\ref{Fig1}c for comparison), is importantly affected by two effects. The singularity in the first term is due to the diverging evaporation profile and has a positive contribution, driving the solute towards the edge. The second term has an opposite effect and depends on the way the radius and the contact angle change with respect to each other. Larger $b$ corresponds to stronger pinning, which reduces the inwards flow contribution and further exacerbates the ``coffee stain" effect. From this, we conclude that the ``coffee stain" effect is driven by two phenomena---evaporation and reduction of the contact line mobility. Evaporation suppression is discussed further in the SI, while the effect of contact line mobility is the main focus of this study.

Parabolic profile approximation and idealized evaporation profile (Equation~\ref{eq:evaporation}) allows to obtain a reasonable approximation for the velocity profile, but the propagation of solute must still be integrated numerically. To model the deposit formation during drying, we used the method of characteristics: integrating trajectories of solute tracers according to the flow profile. Starting from an initial distribution of "tracer molecules" corresponding to a uniform concentration profile, their positions are followed until deposition criteria are met. The particles deposit, when they land outside the droplet radius, or when the local concentration (determined by the current separation of the tracers and the droplet thickness) exceeds the gelling concentration. This leads to the remaining parameter of the model: the ratio of initial and maximal solute concentration ($\phi_0/\phi_\mathrm{max}$). In reality, the gelling transition does not happen abruptly at a maximal solute concentration, but through a continuous increase in viscosity. However, more detailed gelling models for each specific ink would affect the deposited profile in a quantitative way, but would not affect the general features of the deposited profile. 

Let us examine the drying and deposit formation for three possible $R(\theta)$ trajectories: (I.) $b=1$, (II.) $b=0$, (III.) $b=-1/3$, \figurename~\ref{Fig1}b. According to Equation~\ref{RadialFlow}, the radial flow velocity upon drying is substantially higher for pinned drops than for unpinned drops (\figurename~\ref{Fig1}c). The strong radial flow combined with a fixed contact line results in a characteristic ``coffee stain" with nearly all deposit at the periphery (\figurename~\ref{Fig1}d, Drop I). On the other hand, a weaker radial flow in combination with a receding contact line results in much less solute being transferred to the periphery of the drop (\figurename~\ref{Fig1}d, Drop II and III). At $b=0$ (\textit{i.e.,} both radius and contact angle decrease), the drops show only a weak ``coffee stain" deposit, the magnitude of which decreases with the increasing solute concentration (\figurename~\ref{Fig1}d, Case II). The drops with a constant contact angle ($b=-1/3$) exhibit a mountain-like morphology with the solute deposited mostly in the centre of the drop (\figurename~\ref{Fig1}d, Case III). The evolution of droplet shape and accompanying solute deposition upon drying is shown in Supplementary Video 1.

The $R(\theta)$ model neglects backflow and other internal flows, \textit{e.g.,} due to surface tension gradient, viscosity increase, and solute diffusion. We expect these effects to modify the resulting deposition profile in a limited way,\cite{Bodiguel2010} preserving the qualitative morphology imposed by the critical concentration and the $R(\theta)$ trajectory. In addition, the divergence of the evaporation profile at the drop edge is likely suppressed by the accumulation of the less- or non-volatile compounds, including the solute itself. Consequently, the simulated deposition profile will generally overestimate the magnitude of the ``coffee stain'' effect. This effect can be approximated by replacing the evaporation model with a different one without a singularity at the edge of the droplet, which suppresses the ``coffee stain'' effect (see SI Section 2). However, this is merely a heuristic approach, as the complex geometric and chemical processes at the edge do not allow for a simple model derived from the first principles.

\subsection*{2. Methodology and droplet height reconstruction}
\begin{figure*}[t!]
\centering
\includegraphics[clip,angle=0,width=10cm]{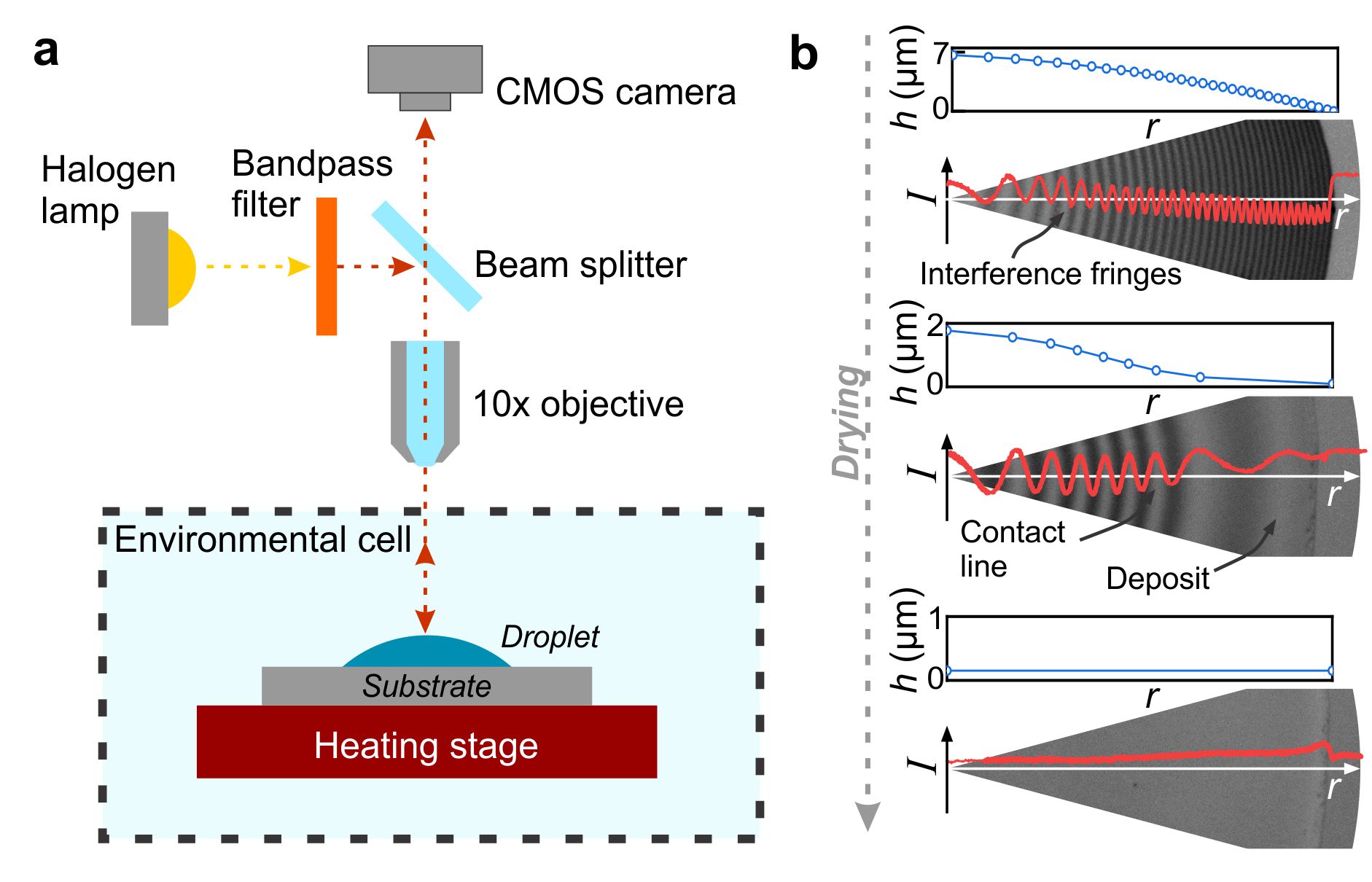}
\caption{\textbf{Experimental setup used to study the drying of the drops.}  \textbf{a,} Schematic of the experimental setup. The drop is illuminated by a narrow bandwidth light and thin-film interference fringes become visible when imaged in reflection mode. \textbf{b,} Snapshots of the same drop at 0~\si{\second} (top), after 50~\si{\second} (middle) and after 70~\si{\second} (bottom) of drying. The drop height profile is reconstructed from the interference image.}
\label{Fig1add}
\end{figure*}

Since the droplets of interest have the thickness in nanometre to micrometre range, we employ a method based on the thin-film interference to reconstruct their height profiles during drying.\cite{Schulze1999} The schematic of the experimental setup is shown in \figurename~\ref{Fig1add}a. The thin-film interference fringes are recorded in a reflection mode by illuminating the drop with a narrow bandwidth light (\figurename~\ref{Fig1add}a). The height profile is then calculated from the interference image. The contact line position and the contact angle at any time can be determined from the derivative of the height profile, $\mathrm{d}h/\mathrm{d}r$ (dotted line in \figurename~\ref{Fig1}a).

To investigate how the contact line motion and evaporation profile impact the shape of dried deposits, nanolitre-sized drops of inks consisting of metal salts dissolved in different solvent mixtures (ethylene glycol- or ethanolamine-based) were printed on surface-modified substrates. A thin polymeric layer was used as a wetting-reset layer, and subsequent UV/ozone treatment was used for surface modification to obtain specific wetting conditions.\cite{Matavz2021} This ensured high reproducibility of the drying experiment and the ability to tailor $R(\theta)$ trajectory for each ink (see Methods for details). 

The following sections describe several drying examples that commonly occur in inkjet printing. First, Section 3 examines the drying of the ethylene glycol-based (EG) ink with pinned contact line. Section 4 discusses the drying of the same ink but with a mobile contact line. Section 5 examines the drying of ethanolamine-based (EA) ink, which produces deposit morphologies very different to EG ink despite similar physical properties of both solvents and $R(\theta)$ trajectories. Every section presents the experimental data and compares it to the predictions of the $R(\theta)$ model. In Section 6, the concepts developed in previous sections are applied to the inkjet-printed square patterns. In contrast to single drops, printing patterns allows to modify the amount of ink per area by adjusting the drop spacing.

\subsection*{3. Pinned contact line}
Constant drop radius (CDR) is the most common mode of drying and has already been extensively studied experimentally and theoretically. Characteristic for the CDR mode is a strong outward liquid flow, which occurs to replenish the solvent evaporated at the edge and maintain the spherical shape of the drop. Consequently, the solute concentrates at the periphery of the drop, which results in the well-known ``coffee stain'' effect.

\begin{figure*}[t!]
\centering
\includegraphics[clip,angle=0,width=14cm]{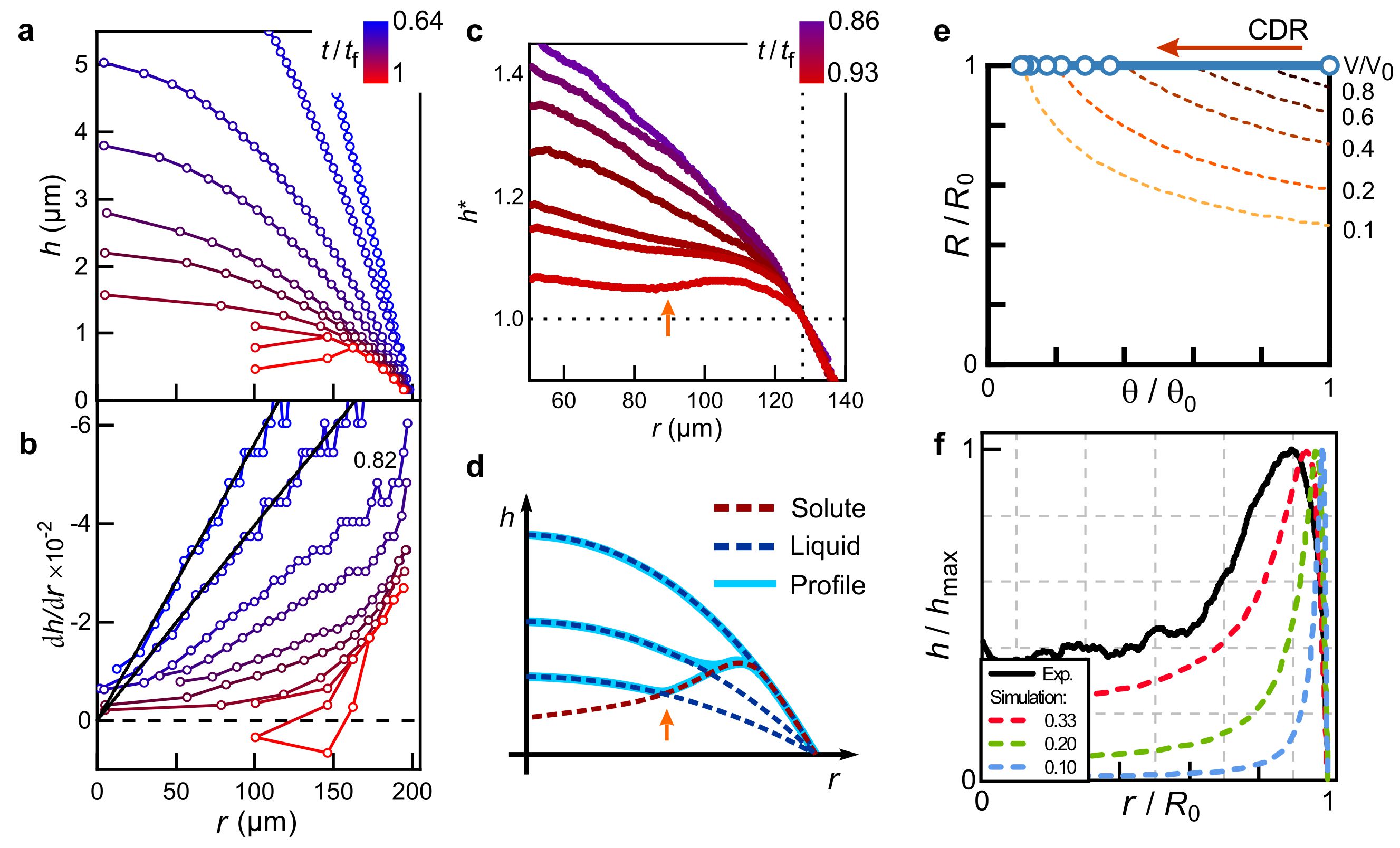}
\caption{\textbf{Drying of the EG ink with a pinned contact line.} \textbf{a,} The height profiles reconstructed from interference images at different drying times and \textbf{b,} the local slope of the height profiles. \textbf{c,} A zoom-in to the region adjacent to the ``coffee stain'' during the last moments of evaporation. Detailed height profiles obtained by the local spectral reflectance measurements are normalised to the height at $r=128$~$\upmu$m for clearer visualisation of the drop shape evolution. The non-normalized height at $r=128$~\si{\micro\meter} for $t/t_f= 0.86$ and 0.93 was 1.7~\si{\micro\meter} and 1.1~\si{\micro\meter}. \textbf{d,} Schematic drawing of a proposed mechanism of height evolution at the very end of drying. The orange arrows in \textbf{c} and \textbf{d} indicate a local thickness depression. \textbf{e,} The $R(\theta)$ phase diagram indicating the CDR mode of drying relative to the initial state ($R_0=200$~\si{\micro\meter}, $\theta_0=16$\si{\degree}). The dotted lines are isochores calculated for a spherical cap geometry. \textbf{f,} Measured thickness profile of dried deposit and prediction of a phenomenological model for three different initial-to-critical solute concentrations.}
\label{Fig2}
\end{figure*}

Here, the CDR mode is exemplified by drops of ethylene glycol-based (EG) ink at the initial contact angle of \SI{16}{\degree}. The initial volume, radius and contact angle were determined using a side-view projection of the drop since it was too thick for interference measurements. The interference fringes became visible at a normalised time, $t/t_f$, of 0.64 ($t_f$ is the total drying time), and the drop at that moment still exhibited a spherical cap geometry (blue data points in \figurename~\ref{Fig2}a). As the drying progressed, the height derivative ($\mathrm{d}h/\mathrm{d}r$, \figurename~\ref{Fig2}b) transitioned from linear to super-linear at $t/t_f>0.82$, which indicates bulging at the drop edge (see SI Section 1 for details). The bulging intensity increased with drying time, and eventually, the drop centre became thinner than its edge, which is evidenced by $\mathrm{d}h/\mathrm{d}r$ changing sign from negative to positive. The final morphology of these drops is a ring-like deposit; hence, the bulging at the edge probably occurs due to the accumulation of the solute and formation of a gelled region.

To provide a greater insight in the last moments of drop evaporation, we reconstructed the droplet height profile using the local spectral reflectance measurements. The detailed height profiles close to the ``coffee stain'' area (\figurename~\ref{Fig2}c) confirm the thickness depression in the centre and suggest that the contact line is permanently pinned and that the drop dries in accordance with the simulations performed by Park \textit{et al.}\cite{ParkS2019} The experiments imply that the formation of the gelled region at the edge does not lead to the formation of a new contact line, but rather the drop smoothly transitions from a spherical cap to a ring as illustrated in \figurename~\ref{Fig2}d. Such drying behaviour describes a typical process of ``coffee stain'' formation, where the super-linear deviation and the centre thickness decrease indicate the solute accumulation at the edge. 
By extracting the drop radius and its contact angle along drying, it is possible to construct the $R(\theta)$ diagram (\figurename~\ref{Fig2}e). Since the drop is always pinned, the $R(\theta)$ trajectory follows the horizontal CDR axis in the diagram. As already discussed in the previous section, pinned drops always produce a ring-like morphology. Comparison of simulated and measured deposit profiles (dashed \textit{vs.} solid lines in \figurename~\ref{Fig2}f) shows that simulations somewhat overestimate the magnitude of the ``coffee stain'', which is expected due to simplifications assumed in our model (see previous section). Nevertheless, the model produces a good qualitative fit to the experimental data.

\subsection*{4. Mobile contact line}
There are infinite possible trajectories in the $R(\theta)$ diagram for drops with a mobile contact line. In the absence of major wetting transitions or out-of-equilibrium initial conditions, the $R(\theta)$ trajectory remains bound in a rectangle with the horizontal line at $R/R_0=1$ (CDR regime) and the vertical line at $\theta/\theta_0=1$ (CDA regime). While the CDR drying regime is common, a purely CDA regime is difficult to achieve due to the contact angle hysteresis. The contact angle hysteresis, which is inherently present due to imperfections of the substrate, causes the contact line pinning in the initial moments of drying and the pinning--unpinning transition upon reaching the receding contact angle. This initial pinning is inevitable for the inkjet-printed drops because they typically exhibit the initial contact angle close to their advancing contact angle due to initial capillary spreading.\cite{Yarin2006}

\begin{figure*}[t!]
\centering
\includegraphics[clip,angle=0,width=12cm]{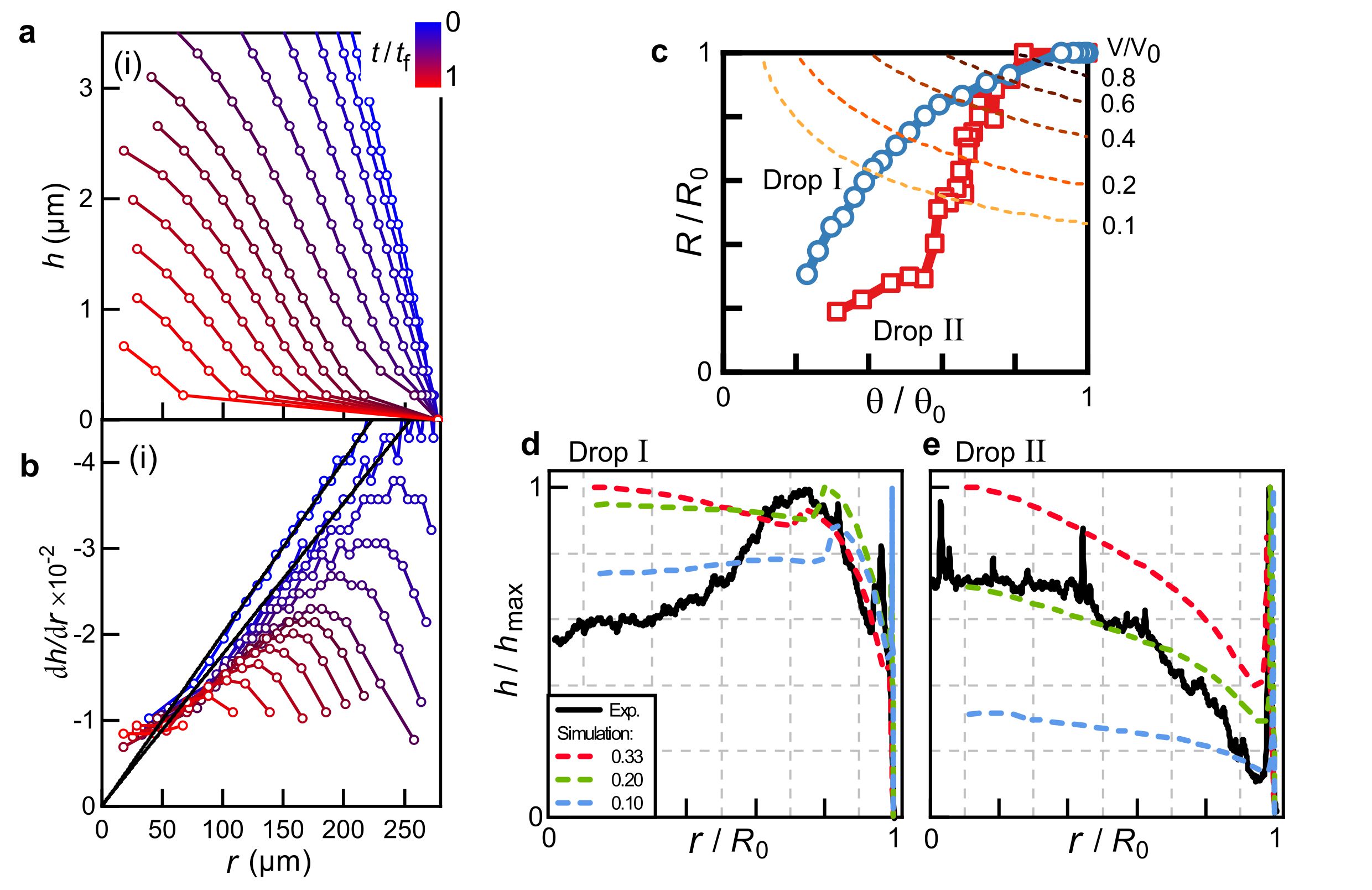}
\caption{\textbf{Drying of the EG ink with a mobile contact line.} \textbf{a,} The height profiles reconstructed from interference images at different drying times and \textbf{b,} the local slope of the height profiles. \textbf{c,} The $R(\theta)$ diagram revealing the initial pinning and the evolution of the contact line for two drops with different trajectories (Drop I: $R_0=320$~\si{\micro\meter}, $\theta_0=5$\si{\degree}, Drop II: $R_0=380$~\si{\micro\meter}, $\theta_0=2.3$\si{\degree}). \textbf{d, e,} The height profiles of dried drops and simulated profiles for three different initial-to-critical solute concentrations for \textbf{d,} Drop I and \textbf{e,} Drop II.}
\label{Fig3}
\end{figure*}

The drying with a mobile contact line is exemplified by the drops of the EG-based ink at the initial contact angle of about \SI{5}{\degree} or smaller. \figurename~\ref{Fig3}a,b show a typical height evolution upon drying of a drop with a mobile contact line. In the first moments of drying, the drop is pinned and exhibits an ideal spherical cap geometry. As the drying proceeds the drop flattens at the edge and a peak appears in $\mathrm{d}h/\mathrm{d}r$ (\textit{e.g.,} the third dataset in \figurename~\ref{Fig3}b). The peak is associated with a formation of an \textit{S}-shaped contact line (SI Section 1), which occurs due to the solute deposition at the receding contact line. The \textit{S}-shaped contact line does not form for the drops without any solute (SI Fig. S2).

For drops with a mobile contact line, the instantaneous position of the contact line ($R$) and respective contact angle ($\theta$) are determined by the local maxima in $\mathrm{d}h/\mathrm{d}r$, which allows simple and straightforward reconstruction of the $R(\theta)$ trajectory from the experimental height profiles. \figurename~\ref{Fig3}c shows the $R(\theta)$ diagram for two drops of EG ink that follow different trajectories upon drying. Both drops are initially pinned; however, Drop I unpins at $V/V_0\approx0.9$ ($\theta/\theta_0=0.93$), while Drop II unpins later at $V/V_0\approx0.8$ ($\theta/\theta_0=0.82$). After the unpinning event, the receding motion of Drop I is characterised by a decrease in both the radius and the contact angle. In comparison, the trajectory of Drop II is much more vertical, indicating a faster decrease in radius compared to the contact angle.

The initial pinning and the $R(\theta)$ trajectory itself produce characteristic features in the deposit morphology (\figurename~\ref{Fig3}d,e). Both drops show a thickness spike at the deposit edge due to the initial pinning, but the spike is much higher for Drop II due to the longer pinning period. Drop I shows a weak ring-like deposit with a height maximum at $r/R_0\approx0.7$. The ring-to-centre ratio ($h/h_\mathrm{max}=0.6$) is much smaller when compared to the pinned drops discussed in the previous section ($h/h_\mathrm{max}=0.3$). On the other hand, Drop II exhibits a dome-like deposit morphology. The dashed coloured lines in \figurename~\ref{Fig3}d,e represent the simulations performed for the $R(\theta)$ trajectories from \figurename~\ref{Fig3}c. The model again slightly overestimates the magnitude of solute transport toward the edge but produces an excellent qualitative fit to the experimental data. The simulations predict the experimentally observed thickness spike at the edge due to the initial pinning and also the deposit's overall shape. For both, Drop I and Drop II, the best agreement between modelled and measured morphology is obtained at the relative concentration $\phi_0/\phi_\mathrm{max} = 0.2$. Supplementary Video 2 shows the simulations of the time evolution of the droplet shape and solute deposition upon drying for Drop II.

\subsection*{5. Mobile contact line with initial peripheral evaporation suppression}
While the previous section demonstrates that the contact-line mobility significantly improves the deposit morphology, some deposition at the edge still occurs due to the initial pinning. This section reports the drying of ethanolamine-based (EA) inks, which were previously reported to produce completely flat deposits in inkjet printing applications.\cite{Matavz2018,Matavz2019} 

\begin{figure*}[t!]
\centering
\includegraphics[clip,angle=0,width=8.5cm]{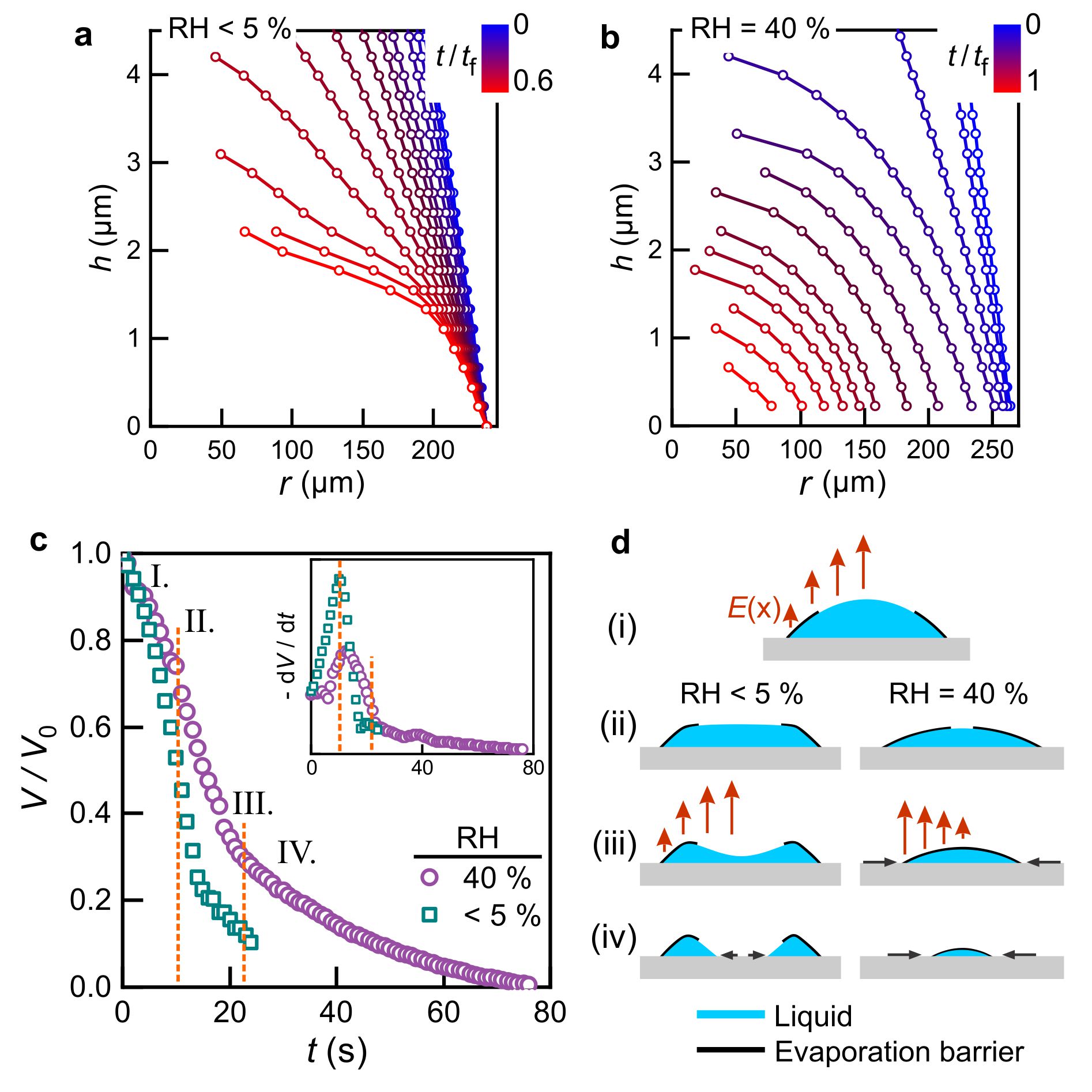}
\caption{\textbf{The drying of EA ink (without solute) at low and moderate relative humidity.} The height profiles reconstructed from interference images at different drying times at \textbf{a,} low relative humidity ($<$5\%) and \textbf{b,} moderate relative humidity ($\sim$40\%). \textbf{c,} The volume evolution as a function of drying time at different relative humidity. \textbf{d,} A schematic of skin formation for drops of EA ink drying at low and moderate RH. Red arrows indicate evaporation intensity and black arrows indicate the contact line motion.}
\label{Fig4}
\end{figure*}

We found that ethanolamine-based (EA) inks exhibit fundamentally different behaviour than EG inks despite very similar physical properties of pure solvents (SI Table S1). While drying of the EG ink is unaffected by the relative humidity, RH (SI Fig.~S3), the humidity strongly affects the drying of EA inks. At low RH ($<\SI{5}{\percent}$), the contact line of EA ink remains pinned and strong bulging at the contact line appears (\figurename~\ref{Fig4}a). At later drying times, $t/t_f>0.6$, the centre of the drop collapses and the drop transforms into a liquid ring (SI Fig. S4). The drying afterwards continues by an outward progression of the newly formed contact line at the drop centre. On the contrary, such behaviour does not occur at moderate RH of $40$\% (\figurename~\ref{Fig4}b), and the drying appears similar to the one observed for the EG ink with a mobile contact line (e.g., \figurename~\ref{Fig3}a). Compared to the EG ink, the EA ink at moderate RH remains pinned much longer (until $t/t_f\approx0.25$) and after un-pinning the contact line recedes with a nearly vertical slope in the $R(\theta)$ diagram (SI Fig. S2).

The analysis of the drop volume evolution with drying time reveals several distinct evaporation regimes for the EA ink (\figurename~\ref{Fig4}c). In the first moments of drying (Region I), the drop volume rapidly decreases and the evaporation rate (inset of \figurename~\ref{Fig4}c) reaches a maximum value in about $10-13$~\si{\second}. This initial increase in the evaporation rate corresponds to ramping up the hotplate temperature (set point reached after \SI{15}{\second}) and also a preferential evaporation of more volatile 2-ethoxyethanol. In the subsequent Region II, the evaporation rate decreases until Point III is reached. At this point, drying at low RH proceeds by the collapse of the drop centre, while at moderate RH the evaporation proceeds normally until \SI{75}{\second}. The overall drying time at low RH is \SI{30}{\second}, which is notably shorter than the drying time at moderate RH. 

\begin{figure*}[t!]
\centering
\includegraphics[clip,angle=0,width=12cm]{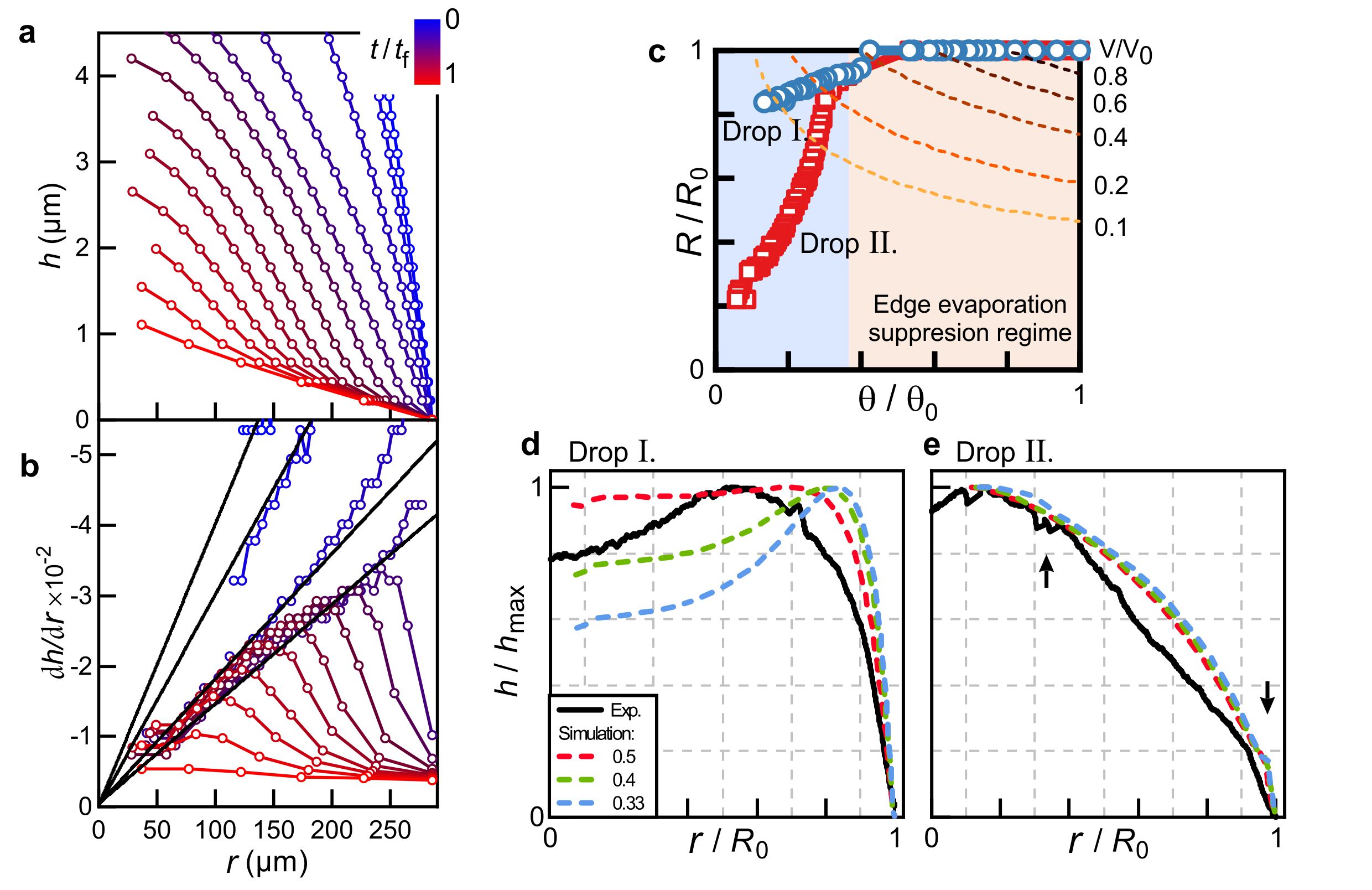}
\caption{\textbf{Drying of the EA ink at moderate relative humidity with a mobile contact line and edge-suppressed evaporation.} \textbf{a,} The height profiles reconstructed from interference images at different drying times and \textbf{b,} the local slope of the height profiles. \textbf{c,} The $R(\theta)$ diagram of two drops with different pinning affinities (Drop I: $R_0=230$~\si{\micro\meter}, $\theta_0=5.8$\si{\degree}, Drop II: $R_0=285$~\si{\micro\meter}, $\theta_0=6.2$\si{\degree}). The orange area indicates the evaporation suppression regime and blue area indicates normal evaporation regime. \textbf{d,}  The height profiles of dried drops and simulated profiles for three different initial-to-critical solute concentrations for \textbf{d,} Drop I and \textbf{e,} Drop II. The black arrows in \textbf{e} indicate the change in the deposit slope, which is due to a momentary change in the $R(\theta)$ trajectory for that drop.}
\label{Fig5}
\end{figure*}
The peculiar drying behaviour of EA-based inks is presumably due to the reactivity of ethanolamine toward ambient CO$_2$ in the presence of water.\cite{Sada1985} The reaction produces N-(2-Hydroxyethyl)-carbamate, which is evident as a dark ring of tiny droplets that condense on the substrate adjacent to the drop periphery (Fig. S5). The formation of carbamate at the drop's liquid--air interface produces skin that retards the evaporation. More skin forms at the drop edge where the mass exchange with the ambient atmosphere is the highest,\cite{DeeganRD_PhysRevE62_2000} and less in the drop centre. This causes the suppression of evaporation at the edge and prevailing evaporation from the centre. At low RH, the skin remains localised to the periphery, which eventually leads to the collapse of the drop centre. In contrast, at moderate RH the skin gradually covers the whole liquid--air interface (Point III in \figurename~\ref{Fig4}c,d). Beyond Point III, the drying at moderate RH is slow and steady, and the height profiles reveal that the drops maintain spherical cap-like geometry (\figurename~\ref{Fig4}b). This implies that the evaporation profile has normalised, and there is no profound evaporation from the centre. Importantly, the contact line becomes mobile before reaching Region IV and recedes at a nearly constant contact angle (SI Fig. S2). The equilibrium reaction of EA with CO$_2$ is reversible;\cite{McCann2009} hence, carbamate decomposes upon evaporation of EA (dark ring in SI Fig. S4 disappears upon drying).

Since the radial flow velocity is dependent on the evaporation profile (Equation~\ref{eq2}), the local suppression of evaporation strongly impacts the solute transport.\cite{Harris2007,DeeganRD_PhysRevE62_2000} To experimentally assess the impact of evaporation suppression on the deposit morphology, the EA-based ink containing solute was deposited on two substrates with different pinning affinities. \figurename~\ref{Fig5}a,b shows that the drop height profiles evolve similarly to the EG ink (\figurename~\ref{Fig3}a,b). The EA ink produces somewhat thicker deposits at the end of drying (profiles shown in red colour), which is consistent with the reactivity theory of EA presented above. Both Drop I and II begin to recede at $V/V_0\approx0.4$; however, the trajectory of Drop I in the $R(\theta)$ diagram is less steep than for Drop II (\figurename~\ref{Fig5}c). The difference in trajectory slope is clearly reflected in the deposit shape, whereas Drop I exhibits a weak ``coffee stain'' and Drop II exhibits a dome-shaped deposit. Both profiles exhibit no deposit spike at the edge (as in the case of EG ink), which is attributed to the evaporation suppression at the edge due to the skin formation. 

To account for the non-trivial behaviour of EA ink in the simulations, only the blue region in the $R(\theta)$ diagram (\figurename~\ref{Fig5}c) was considered in the drying simulations. The region was selected according to \figurename~\ref{Fig4}c, which shows that the evaporation profile stabilises after $V/V_0 \approx 0.35$ (Point IV). About 65\% of solvents already evaporated at this point and the initial concentration of solute in both drops is accordingly higher. This is a very approximate approach since the temporal evolution of the evaporation profile is unknown. Nevertheless, the modelled deposit morphology for both drops (dashed lines in \figurename~\ref{Fig5}d,e) agree well with the experimental data (solid black lines). Despite very complex behaviour of the EA ink, the model reproduces even small details in the deposit morphology, such as the change in deposit slope due to changes in the $R(\theta)$ trajectory (indicated by the black arrows in \figurename~\ref{Fig5}e) .

\subsection*{6. Drying of inkjet-printed features}
\begin{figure*}[t!]
\centering
\includegraphics[clip,angle=0,width=8cm]{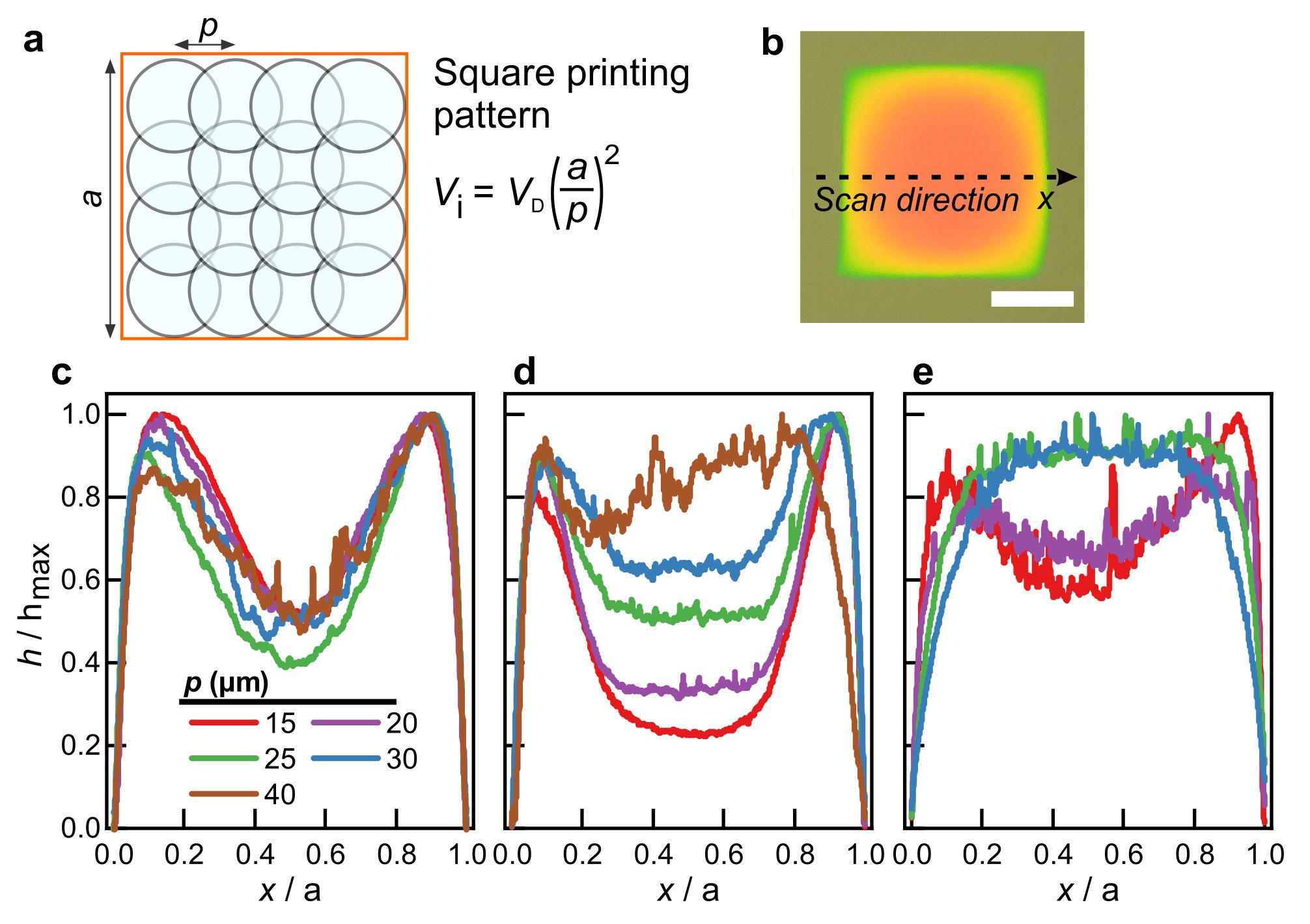}
\caption{\textbf{The deposit morphology of a square pattern printed at different drop spacings.} \textbf{a,} Schematic of the printing process of square pattern. Adjacent drops are separated by a distance called drop spacing, $p$. \textbf{b,} Optical micrograph of printed and dried square pattern from EA/EG ink. Colours reflect the deposit thickness profile, where the thickness increases from green, yellow to orange. The normalised thickness profile of squares printed using the \textbf{c,} EG ink with a pinned contact line, \textbf{d,} EG ink with a mobile contact line, \textbf{e,} EA/EG ink with a mobile contact line and initial suppression of the evaporation at the edge.}
\label{Fig6}
\end{figure*}

To investigate how the concepts developed above reflect in the technologically relevant printing applications, we reproduced three drying scenarios for the inkjet-printed square pattern. In contrast to single drops discussed so far, the printing of a square pattern allows tailoring of the ink volume-per-area ratio by adjusting the pitch between the adjacent drops, \textit{i.e.,} the drop spacing $p$ (\figurename~\ref{Fig6}a). Hence, the drop spacing sets the initial contact angle of the printed feature (within the limits of the contact angle hysteresis) and, consequently, the period of initial pinning in the systems that undergo a pinning-unpinning transition during drying. 
The squares printed from the EG ink with a pinned contact line exhibit the deposit morphology that is unaffected by the drop spacing (\figurename~\ref{Fig6}c). On the other hand, drop spacing strongly impacts the morphology of the features that dry with a mobile contact line. A transition from ring-like to flat or slightly dome-shaped deposit is observed for the EG ink and the EA/EG ink with a mobile contact line (\figurename~\ref{Fig6}d,e). EA/EG ink is a ternary solvent ink composed of 2-ethoxyethanol, EA and EG in 20:64:16 volume ratio---the composition that previously proved efficient for deposition of high-quality metal-oxide structures.\cite{Matavz2018,Matavz2019} Adjusting the ratio between EG and EA allows tailoring of the evaporation profile and to essentially obtain completely flat deposits at reasonable ink concentrations and drop spacing values. The evolution of morphology with the drop spacing is consistent with our phenomenological model: increasing the drop spacing shortens the period of initial pinning and results in less solute deposited at the edge. As a consequence, the patterns printed from inks with a mobile contact line produce deposits with "rounded" corners (\figurename~\ref{Fig6}b), where the thickness less gradually increases than in other areas. This occurs due to intrinsically smaller contact angle at the edge of a square pattern.\cite{Soltman2013} In contrast, the patterns drying with pinned contact line exhibit bumps at the corners due to stronger capillary flow to corners to replenish the evaporated solvent.

In the scope of the $R(\theta)$ model it is possible to discuss the impact of other experimental conditions on the deposit formation. The first such condition is the solute itself. We found that solutes with a higher pinning affinity (\textit{e.g.,} transition metal alkoxides) produce deposits with a more pronounced edge deposit than the solutes with a lower pinning affinity (\textit{e.g.,} metal salts), SI Fig. S5.\cite{phdthesis} The solutes with a higher pinning affinity shift the $R(\theta)$ trajectory toward longer pinning times, which results in more solute transport to the edge. The drying temperature can also affect the deposit morphology by changing the pinning affinity. Since there is no general relation between the substrate temperature and the contact angle, the temperature will likely affect the morphology of deposits of certain ink--substrate combinations differently.\cite{Robin2015,Petke1969} In our study the temperature had a subtle effect: higher temperatures (in the range from 40~\si{\celsius} to 80~\si{\celsius}) produced slightly stronger edge deposit.\cite{phdthesis}

The last point of discussion is the role of the surface-tension-driven flow (Marangoni flow) on the deposit formation of solution-based inks. The Marangoni flow may have a significant role in (thick) drops of particle-dispersion inks for which the relevant length scale depends on the particle size.\cite{Monteux2011} However, many studies reveal scenarios that cannot be explained in the scope of the Marangoni flow. Three representative examples are: (i) Babatunde et al. investigated the effect of solvent composition on the internal liquid flows and final deposit morphology. No general relation between flow characteristics and deposit morphology.\cite{Babatunde2013} (ii) Eagles et al. performed a study on modelling of thin, binary liquid droplet drying with pinned contact line. The study revealed that flat or dome-shaped deposits cannot be reproduced at realistic drying condition.\cite{Eagles2016} (iii) Diddens et al. report detailed modelling of evaporation of multi-component droplets.\cite{Diddens2017} They demonstrated the preferential evaporation of a more volatile solvent and showed that the surface-tension-driven flow typically vanishes before the drying is complete. In addition to these three examples, the influence of experimental parameters, such as temperature, drop spacing, solvent/solute effects and wettability, cannot be adequately described by considering the Marangoni flow solely. Although this work does not explicitly investigate the role of Marangoni flow in drying drops, it does imply its minor role in deposit formation. Instead, we demonstrate that contact-line mobility plays a significant role in deposit morphology, while other variables account for the minor differences between the predicted and measured deposit morphology.

\section*{Conclusions}

This work highlights the role of contact-line mobility in the deposit formation in solution-based inks at lateral dimensions and contact angles relevant for printing applications. The developed phenomenological model uses the input parameters that can be readily extracted from experimental data. This allows a direct comparison of modelled and measured deposit shapes and represents a powerful improvement over previous models that employ the input parameters that are difficult to assess experimentally.\cite{Doi2016,Frastia2011,Brown2014} We show that a mobile contact line is crucial for overcoming the ``coffee stain'' effect. The deposit formation was greatly affected by the $R(\theta)$ trajectory and evaporation profile during drying. The pinning of the contact line during the initial moments of drying produced a spike in deposit thickness at the edge for the drops with normal (divergent) evaporation profile. We demonstrated that thickness spike can be suppressed by using the solvent that reacts with ambient air and forms a ``skin'' that locally suppresses the evaporation. The concepts learned on single drops were translated to inkjet-printed patterns that consist of many drops separated by the drop spacing. The patterns that dry with a pinned contact line exhibit an identical deposit morphology regardless of the drop spacing value. On the other hand, the patterns with mobile contact line show a strong dependence of the deposit morphology on the drop spacing value, which is consistent with the proposed $R(\theta)$ model.

The contact line motion occurs at very low contact angles, which cannot be resolved by the conventional side-view drop observation that was used in most of the drying studies on the influence of the internal liquid flows on the deposit morphology..\cite{Hu2006,Babatunde2013} Hence, previous studies that reported deposit morphologies other than ``coffee stain'' might overlooked the contact line motion and instead overestimated the importance of Marangoni flow on deposit shape.

\section{Acknowledgement}
This work was financed by Slovenian Research Agency under programs P1-0125, P2-0105, P1-0099, and projects J1-9149, J2-1740, N1-0104;

\bibliographystyle{elsarticle-num}

\end{document}